\documentclass[prb,aps]{revtex4}
\usepackage{color,graphicx,bm,amsmath}

\newcommand{\eq}[1]{Eq.(\ref{eq:#1})}

\newcommand{\half}{\ensuremath{\frac{1}{2}}}
\newcommand{\halfs}{\mbox{$\frac{1}{2}$}}

\newcommand{\degree}{\ensuremath{^{\circ}}}

\begin{document}

\pacs{42.65.Yj,78.20.Bh,42.60.Jf}

\title
{Numerical Modelling of the Microcavity OPO}

\author{D. M. Whittaker}

\affiliation{Department of Physics, University of Sheffield, S3 7RH, Sheffield, United Kingdom}

\begin{abstract}
A detailed description is provided of a numerical model of the microcavity
optical parametric oscillator (OPO). The model is solved on a two
dimensional grid, in the time domain, using an alternating-direction
implicit method. Results are presented for a system pumped  with a
gaussian spatial profile beam. It is shown that at high pump powers
the OPO signal forms a ring, with low emission at the centre where the
pumping is strongest.\\
\begin{center}
To be published in Physica Status Solidi (c)
\end{center}
\end{abstract}

\maketitle                   % Produces the title.

\section{Introduction}

The polariton physics in a strongly-coupled semiconductor microcavity
includes interesting non-linear optical effects. When resonantly
pumped with an off-axis laser, optical parametric oscillator (OPO)
behaviour can be obtained, with the spontaneous appearence of signal
and idler emission\cite{stevenson2000,baumberg2000}. Since the
excitonic non-linearity is $\chi^{(3)}$, the phase matching conditon
requires $2 \omega_p = \omega_s + \omega_i$ and $2 k_p = k_s + k_i$,
which means that the pump, signal and idler can all be close to
resonance with the polariton dispersion. This behaviour has been
investigated theoretically by a number of authors, using
analytic\cite{ciuti2000,whittaker2001} and numerical\cite{gippius}
models. However, all these models differ from the experimental setup
by assuming that the pump is a plane wave, with uniform excitation of
the system, and uniform polariton fields. While such treatments
provide useful insights into the OPO, a fuller understanding of the
observed behaviour requires a theory which takes into account the
finite excitation spot used in experiments. The purpose of this paper
is to describe a numerical treatment of a two dimensional microcavity
polariton system pumped with a Gaussian beam, solved in the time
domain on a discrete spatial grid.

\section{The Numerical Model}

\newcommand{\hop}{H^{\rm hop}}
\newcommand{\site}{H^{\rm site}}
\newcommand{\mx}{\ensuremath{\mbox{\tiny{\rm X}}}}
\newcommand{\mc}{\ensuremath{\mbox{\tiny{\rm C}}}}
\newcommand{\tx}{t^x_{\mx}}
\newcommand{\tc}{t^x_{\mc}}
\newcommand{\px}{\phi_{\mx}}
\newcommand{\pc}{\phi_{\mc}}

The model for the OPO is based on the non-linear optics treatment of
Ref.\cite{whittaker2001}. It consists of coupled two-dimensional
cavity and exciton fields, $\pc$ and $\px$, with a local $\chi^{(3)}$
non-linearity in the exciton to represent the effects of short-range
interaction. Both fields have dispersions, described by effective
masses $M_{\mc}$ and $M_{\mx}$, and are damped with widths
$\gamma_{\mc}$ and $\gamma_{\mx}$. The coupling is characterised by
$\Omega$, the Rabi splitting, and $\Delta$, the detuning at normal
incidence.  The cavity field is driven by a spatially dependent
external pump field $f_p(r) \exp{(i \omega_p t)}$. The time evolution
of the fields is thus determined by
\begin{eqnarray}
i \frac{\partial \pc}{\partial t} &=&
    \left(-\frac{1}{2 M_{\mc}} \nabla^2 -i \gamma_{\mc}  + \Delta \right) \pc
        + \half \Omega \, \px + f_p(r) \exp{(i \omega_p t)} \\
i \frac{\partial \px}{\partial t} &=&
    \left(-\frac{1}{2 M_{\mx}} \nabla^2 -i \gamma_{\mx} + \kappa |\px|^2 \right) 
     \px + \half \Omega \, \pc
\end{eqnarray}

To be exact, the form for the $\chi^{(3)}$ nonlinearity should be
$\kappa \px^3$. However, for a numerical treatment, it is better to
chose $\kappa |\px|^2 \px$, as this is independent of the origin of
energy, which can thus be shifted to the exciton frequency. The
modified form correctly gives the scattering effects of
interest here, but does not describe frequency multiplying effects,
which would require extremely small time-steps in the numerics.

For the numerical treatment, this model is discretized on a two dimensional
grid, with the differential operators are replaced by hopping
terms. The solution is obtained in the time domain, as an initial
value problem, using the Crank-Nicholson method\cite{crank}. This leads to a
semi-implicit time-step formula of the form
\begin{equation}
 \left( 1 + \halfs i H \delta t \right) \psi^{n+1}_{ij}=
 \left( 1 - \halfs i H \delta t \right) \psi^{n}_{ij},
 \label{eq:crank}
\end{equation}
where the vector $\psi^n_{ij}=(\pc,\px)^T$ describes the
cavity photon and exciton amplitudes on the site $(ij)$ at time-step $n$.
The precise form of the matrix operator $H$ will be specified below.

The problem with solving \eq{crank} in two dimensions is that $H$ is a
large matrix (size $N^2 \times N^2$, where N is the length of the edge
of the grid), whose structure does not lend itself to efficient
numerical solution: there are non-zero terms far from the
diagonals. It is better to use an operator-splitting treatment, the
alternating direction implicit method\cite{splitting}, writing $H=H_x
+ H_y$, where $H_x$ and $H_y$ respectively contain only $x$ and $y$
hopping terms. Then, each time-step is divided into two sub-steps:
first,
\begin{equation}
 \left( 1 + \halfs i H_x \delta t \right) \psi^{n+\half}_{i(j)}=
 \left( 1 - \halfs i H_x \delta t \right) \psi^{n}_{i(j)}
 \label{eq:xup}
\end{equation}
is solved for every $j$, then 
\begin{equation}
 \left( 1 + \halfs i H_y \delta t \right) \psi^{n+1}_{(i)j}=
 \left( 1 - \halfs i H_y \delta t \right) \psi^{n+\half}_{(i)j}
 \label{eq:yup}
\end{equation}
for every $i$. Each of these one dimensional strip problems is independent,
so it is necessary to solve only $2N$ $(N \times N)$ sets of
equations. Moreover the matrices $H_x$ and $H_y$ contain non-zero
terms only on the diagonal and first three sub-diagonals, which permits a
very efficient solution by factorisation\cite{details}. 

The part of $H_x$ involving coupling to site $i$ is
\begin{equation}
 H_x =
 \left(
   \begin{array}{ccccccc}
    0& \hop & \half \site_{i-1}  & \hop   &   0  & \cdots & \cdots \\
     \cdots &0& \hop & \half \site_{i}  & \hop   & 0  & \cdots \\
     \cdots &  \cdots & 0 & \hop& \half \site_{i+1}  & \hop & 0  \\
   \end{array}
 \right)
\label{eq:strip}
\end{equation}
where each $H$ is a $2 \times 2$ matrix:
\begin{equation}
 \hop=
  \left(
    \begin{array}{cc}
	-\tc & 0 \\
         0  & -\tx
     \end{array}
  \right)
 \; \;
 \site=
  \left(
    \begin{array}{cc}
	2\tc - i \gamma_{\mc} + \Delta &  \halfs  \Omega \\
         \halfs  \Omega  & 	2\tx - i \gamma_{\mx} + \kappa |\px|^2  \\
     \end{array}
  \right).
\end{equation}
The hopping terms are $\tc=(2M_{\mc} \delta_x^2)^{-1}$ and
$\tx=(2M_{\mx} \delta_x^2)^{-1}$, with $\delta_x$ the $x$-direction
grid spacing. The value of $\px$ used in the non-linear term is the old
value at that site, ie from the right hand side of \eq{xup} or
\eq{yup}.  $H_y$ has a similar form, except $\delta_y$ replaces
$\delta_x$, though in practice an isotropic grid is used. The factors
of \half\/ multiplying $\site$ are needed to avoid over-counting, because
they occur in both $H_x$ and $H_y$.

\eq{xup}, and similarly \eq{yup}, is most efficently solved in two
stages: first an intermediate variable $\psi^t$ is calculated using
\begin{equation}
\half  \left( 1 + \halfs i H_x \delta t \right) \psi^t = \psi^{n},
\label{eq:trick}
\end{equation}
then it is straightforward to show that $\psi^{n+\half}=\psi^t - \psi^n$.

To incorporate periodic boundary conditions, 
\eq{strip} has to be modified by adding hopping terms in the
off-diagonal corners of the matrix, coupling the sites at either end
of a strip:
\begin{equation}
  H_x^p=
  \left(
   \begin{array}{ccccccc}
       &  & & & \cdots & 0 & \hop \\
       &  &  & & \cdots & 0 & 0 \\
      \vdots  & \vdots & & & & \vdots & \vdots  \\
      0 & 0 & \cdots & & &  &  \\
     \hop & 0 & \cdots & & &  & 
   \end{array}
  \right).
  \label{eq:periodic}
\end{equation}

Adding these terms breaks the useful diagonal form of the matrix, so a
direct solution of \eq{periodic} would be inefficient. However, as there are
just four extra non-zero entries, the problem can be handled at little
numerical cost using the Woodbury formula\cite{woodbury} to incorporate the extra
terms. $H_x^p$ is written as
\begin{equation}
 H_x^p = H_x' + \sum_{k=1,2} u_k \otimes v_k 
\end{equation}
where the $u_k$ and $v_k$ are chosen to generate the end hopping
terms. An efficent choice is
\begin{eqnarray}
u_1&=&( \beta_1, 0, \ldots , -\tc,0)^T\\
v_1&=&(1, 0, \ldots , -\tc/\beta_1,0)^T \\
u_2&=&(0, \beta_2, \ldots , 0, -\tx)^T\\
v_2&=&(0, 1, \ldots , 0, -\tx/\beta_2)^T 
\end{eqnarray}
where $-\beta_1$ and $-\beta_2$ are the first two elements along the
main diagonal of $H_x$. $H_x'$ is a slightly modified version of
$H_x$: the $u \otimes v$ terms produce, as well as the required
hopping terms, four additional elements along the main diagonal which
have to be subtracted from $H_x$ to obtain the correct $H_x^p$.    

Evaluating the Woodbury formula requires additional solutions of
\eq{trick} with $u_1$ and $u_2$ on the right hand side, but this is
very inexpensive once the factorisation of $H_x'$ is obtained. Hence
adding periodic boundary conditions makes little difference to the
numerical costs.

\section{Results}

The results described in this section are for a microcavity with
physical parameters typical of a good GaAs structure: the Rabi
splitting, $\Omega=5$ meV, and the line-widths of the exciton and
cavity mode are $\gamma_{\mx}=\gamma_{\mc}=0.25$ meV. The structure is
on-resonance at normal incidence ($k=0$), and pumped at 16\degree\/ with
a Gaussian beam whose FWHM is 17.05$\mu$m. The cavity mode effective mass
is taken to be $3 \times 10^{-5} m_e$, and the exciton mass $0.5
m_e$. The spatial grid consists of $(2^7 \times 2^7)$ points with a
separation of 0.4$\mu$m, and the time step is 12 fs.

The calculation is started with the fields set to zero, and the
solution is then propagated until a steady state is reached, typically
after about $10^4$ time steps or $\sim 100$ ps. The pump, signal and
idler are then separated by storing the data for a series of time
steps, Fourier transforming to the frequency domain, and applying
band-pass filters. This process can be carried out in a continuous way
using a rolling time window, which allows the observation of dynamical
effects occuring on time-scales slow compared to the inverse of the
mode separation.

The calculated images show the spatial dependence of the steady-state
cavity field intensities in the signal and idler modes at various pump
powers.  It should be noted that they represent the internal cavity
field, which cannot be directly observed experimentally. The emitted
light will undergo some additional spatial filtering due to the small
$k$-width of the polariton branches.

\begin{figure}[t]
\begin{center}
\includegraphics[scale=.5]{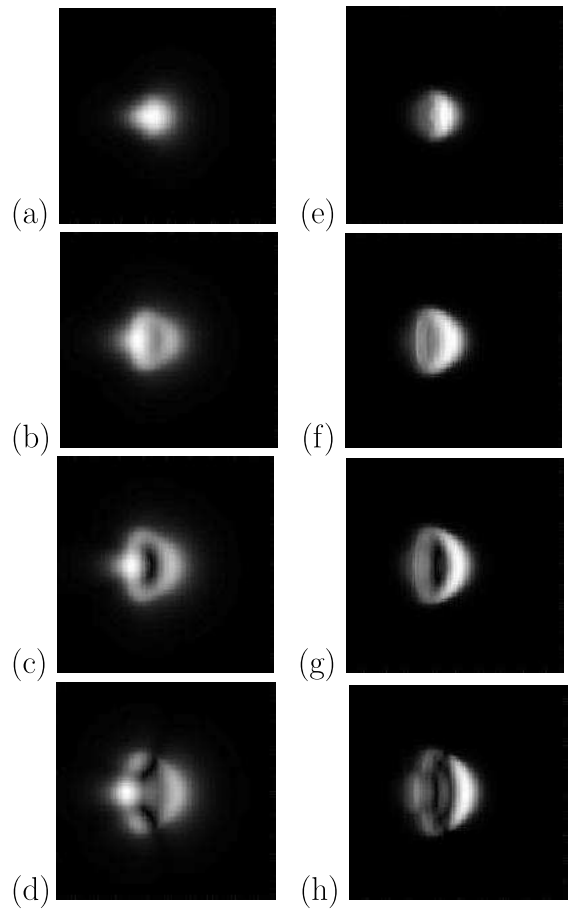}
\end{center}
\caption{Grey-scale plots of the OPO states for Gaussian pump beam at
16\degree with various powers. (a) and (e) are the signal and idler
intensities for a peak power of 0.2, just above threshold. The
following rows represent pump powers of 0.3, 0.4 and 0.5, with the
signal states on the left, the idlers on the right. The lengths of the
edges of the squares are 51.2$\mu$m, and the FWHM of the pump beam is
17.05$\mu$m.
}
\end{figure}

Fig.1 shows the calculated signal and idler images for the steady
state at various pump powers above threshold. Close to threshold, (a)
and (e), only a small spot near to the centre of the image is
bright. This is easily understood: for a Gaussian beam profile, the
threshold will be reached first where the pump is strongest.  At
higher pump powers, the emission spot expands, but, instead of the
centre becoming brighter, the spot turns into a ring with low
intensity in the middle. It can be explained qualitatively using an
analytic plane-wave model where the contributions of the signal and
idler to the polariton blue-shifts are taken into
account\cite{whittaker2004}. The shifts cause the OPO to switch off
above a certain power, which depends on the incidence angle and
detuning of the pump beam. This happens first at the centre of the
spot, leading to the dark middle. In a Gaussian beam, there will
always be a region where the pump power is the range to allow the OPO,
so an expanding emission ring should be expected as the power is
increased above threshold.

Of course, the strong spatial dispersion of the polariton
means that a such a simple picture, relating the OPO intensity to the local
value of the pump, is not completely accurate, but it seems to give a
reasonable guide. The dispersion effects are responsible for the fact
that the ring is not perfectly symmetrical, and breaks up
significantly at higher pump powers. However, a bilateral symmetry is
always mainatined, defined by the incidence plane, which cuts the
figure along the $x$-axis.

Finally, at higher powers than those shown in the Fig.1, the numerics
predict that a steady state may not be reached. Instead, the signal
and idler undergo a periodic oscillation, maintaining the general ring
shape, but with the maximum intensity moving from side to side.

\section{Summary}

It has been shown that a full understanding of the microcavity OPO requires
a two-dimensional treatment of the spatial structure of the pump beam and
the polariton fields. A numerical model describing the spatial and
temporal evolution of the system has been derived, and used to predict
that the emission spot should form a ring at high pump powers. These
results demonstrate that experimental measurements without spatial
resolution provide only a limited picture of the OPO behaviour.

\end{document}